\documentclass[12pt,fleqn]{article}
\usepackage{amssymb}
\newcommand{\vek}[1]{\mbox{\bf #1}}
\newcommand{\be}{\begin{equation}}
\newcommand{\ee}{\end{equation}}

\begin{document}
%\hspace*{\fill} LMU-TPW-99/24\\[2ex]
\begin{center}
\Large\bf
Gravity and the Newtonian limit in the Randall--Sundrum model
\end{center}
\vspace{2ex}
%\section*{}
\normalsize \rm
\begin{center}
 R. Dick, D. Mikulovicz\\[0.5ex] 
{\small\it
Sektion Physik der Ludwig--Maximilians--Universit\"at\\
Theresienstr. 37, 80333 M\"unchen, Germany}
\end{center}

\vspace{5ex}
\noindent
{\bf Abstract:} 
We point out that the gravitational
evolution equations in the Randall--Sundrum model
appear in a different form than hitherto assumed. As
a consequence, the model yields a correct Newtonian limit
 in a novel manner.

\newpage  
\section{Introduction}

Randall and Sundrum recently explained the large
hierarchy between the Planck scale and the weak
scale in terms of a model where our observable
universe corresponds to a $(3+1)$-dimensional boundary of
a $(4+1)$-dimensional manifold. The extra spacelike dimension
in this model
is non-periodic and finite \cite{RS1} or eventually even of infinite
length \cite{RS2}, and ordinary matter is restricted to the
boundary while gravitational modes may propagate in the bulk.
This model attracted a lot of attention, see \cite{more}
and references there.

Usually the Randall--Sundrum model is investigated under the
assumption that the coupling of gravitational modes to
the matter on the boundary is governed by the restriction
of the Einstein tensor to the boundary. This led in particular
to the claim
that the Randall--Sundrum model
with a small non-periodic dimension would yield antigravity
\cite{SMS}.

In the present paper we reconsider 
the gravitational evolution equations and the Newtonian limit
of the Randall--Sundrum model. Contrary to the common assumption,
the coupling of matter to gravity in this model does not appear
through an Einstein equation on the boundary but through Neumann
type boundary conditions for the Einstein equation in the bulk.
This deviation from Einstein gravity
on the boundary cures the antigravity problem:
Gravity on the boundary is attractive and has a correct Newtonian
limit.

The observation that the Randall--Sundrum model implies first order
equations for the metric on the boundary instead of second order
equations does not depend on whether one uses the Einstein--Hilbert
term or an Einstein term for the gravitational action in the bulk.
However, with the Einstein--Hilbert term the system of gravitational
evolution equations is overdetermined and therefore we use an Einstein
term. To explain this, we point out several differences
between the Randall--Sundrum model and Kaluza--Klein theory
in the next section before we address the equations
of motion for the metric and the Newtonian limit
in the Randall--Sundrum model.

To avoid confusion,
we count the dimension of spaces with Minkowski
signature explicitly in the form $d+1$, with $d=3,4$. 

\section{Differences to Kaluza--Klein theory}

One might presume that gravity should contribute an
Einstein--Hilbert term to the action of the Randall--Sundrum model,
and that 
 gravity in $3+1$ dimensions should arise 
in a similar way as in a Kaluza--Klein
theory with periodic dimensions.

In
Kaluza--Klein theories with small
periodic internal dimensions the low-energy degrees of freedom are restricted
to zero modes which are separated by a large mass gap from
the massive modes. 
The zero modes are indepedent from the internal coordinates
and the resulting low-energy theory is genuine $(3+1)$-dimensional.
However,
the periodicity constraints are instrumental
for the emergence of the mass gap. Contrary
to Kaluza--Klein theory, boundary conditions
in a bounded {\sl non-periodic} dimension have to be fixed
dynamically by the equations of motion, and {\it a priori} 
this does not imply
a restriction to zero modes separated by a mass gap.
By the same token, we have to subtract a complete divergence
from the Einstein--Hilbert term in the action of the
Randall--Sundrum model.

 For an explanation of this point, consider
the Einstein--Hilbert action with a cosmological term
in the $(4+1)$-dimensional universe
of the Randall--Sundrum model ($d^5x=d^4xdx^5$, $0\le x^5\le L$):
\[
S_{EH}=\int d^5x\sqrt{-g}\Big(\frac{\mu^3}{2}g^{MN}R_{MN}-\Lambda\Big),
\]
\be\label{dseh}
\delta S_{EH}=\int d^5x\sqrt{-g}\delta g^{MN}
\Bigg(\frac{\mu^3}{2}
\Big(R_{MN}-\frac{1}{2}g_{MN}g^{KL}R_{KL}\Big)
+\frac{1}{2}g_{MN}\Lambda\Bigg)
\ee
\[
+\frac{\mu^3}{2}\oint d^4x\sqrt{-g}\Big(g^{MN}\delta\Gamma^5{}_{MN}
-g^{5N}\delta\Gamma^M{}_{MN}\Big)\Big|_{x^5=0}^L.
\]
If matter degrees of freedom could propagate on the
whole manifold and if
the fields would be periodic, then\\
$\bullet$ the boundary terms would cancel\\
and\\
$\bullet$ we could perform a Fourier decomposition
of the degrees of freedom and throw away the massive Kaluza--Klein
modes.

This would then correspond to original Kaluza--Klein
theory and yield low-dimensional Einstein gravity
in the usual way.
However, the space-time points $x^0,\ldots x^3,x^5=0$
and $x^0,\ldots x^3,x^5=L$ are different physical points in 
the Randall--Sundrum model
and periodicity is not required (and cannot be required
by causality). Furthermore, matter degrees of freedom are supposed
to be fixed to the $(3+1)$-dimensional boundaries, and therefore
variation of corresponding action principles yields
homogeneous equations for the gravitational degrees
of freedom in the bulk, while the coupling to the matter
degrees of freedom arises from the variation on the boundaries.
Below we will point out that the gravitational
potential in this theory does not
correspond to a three-dimensional Greens function for 
Dirichlet boundary conditions
at infinity, but to a four-dimensional 
Greens function for Neumann boundary conditions
on three-dimensional boundaries. 
{\it A priori} this implies deviations from the ordinary
Newton potential in three dimensions. 
However, for a small non-periodic extra dimension 
of length $L$ the
leading terms in the Newtonian limits of the Randall--Sundrum
model and
Einstein gravity agree if the naive estimate
on the relation between the $(3+1)$- and $(4+1)$-dimensional
 Planck masses (inferred from the corresponding
relation in Kaluza--Klein theory)
is augmented by a factor 3. Another difference 
to Kaluza--Klein theory concerns the fact, that 
for a small non-periodic extra dimension
 the deviations from the Newtonian
limit of Einstein gravity are not suppressed by a term
 $\exp(-r/L)$ but correspond to an expansion in
 $r/V_3^{1/3}\ll 1$, where $V_3$
is the 3-volume of a time slice
of the $(3+1)$-dimensional boundary. 
As a consequence, in the large $V_3$ limit
the gravitational potential has the usual form, with the correction
term corresponding to a renormalization
of the Planck mass.

\section{The gravitational potential in the Randall--Sun\-drum model}

We have seen that
the excitation of space-time curvature
in the Randall--Sundrum model
does not arise due to
matter sources in the bulk equations, but through boundary conditions
on the gravitational field arising from boundary equations
of motion. This raises the issue of the Newtonian limit
for the gravitational field
on the boundary, which we examine through the $(4+1)$-dimensional
action
\be\label{action}
S=\int_{\cal V} d^5x \sqrt{-g}\Bigg(\frac{\mu^3}{2}g^{KL}
(\Gamma^M{}_{NK}\Gamma^N{}_{ML}-\Gamma^M{}_{NM}\Gamma^N{}_{KL})-\Lambda\Bigg)
+\sum_{i=1}^2\int_{\partial{\cal V}_i} d^4x{\cal L}_i.
\ee

Here $\partial{\cal V}_i$ are the
two connected components of the $(3+1)$-dimensional boundary 
and ${\cal L}_i$ denotes
 the Lagrangians for the matter degrees of freedom
on the boundary components.
 The Lagrangians ${\cal L}_i$ may also contain cosmological terms on the boundary.

Coordinates $x^0,x^1,x^2,x^3,x^5$ are chosen such that the two
boundary components correspond to $x^5=0$ and $x^5=L$, respectively.

The gravitational part in (\ref{action}) is fixed from two requirements:\\
$\bullet$ The Einstein tensor is the leading derivative term
in any evolution equation for the metric on a Riemannian
manifold, and this should also hold true in the present model, 
since there is no symmetry prohibiting this leading
curvature term.\\
$\bullet$ At the same time, the full Einstein--Hilbert
Lagrangian (as well as a leading higher curvature term
in the action) would not give consistent boundary equations of
motion, due to the second
derivatives on the metric tensor:

The divergence
included in the Einstein--Hilbert
Lagrangian yields boundary terms $\sim\partial_5\delta g_{\mu\nu}$
which have no counterpart in the $\delta{\cal L}_i$ terms and
overdetermine the boundary value problem for the metric.
Therefore, we used Einstein's well-known
Lagrangian (adapted to $4+1$ dimensions)
\[
{\cal L}_E=\frac{\mu^{3}}{2}\sqrt{-g}g^{KL}
(\Gamma^M{}_{NK}\Gamma^N{}_{ML}-\Gamma^M{}_{NM}\Gamma^N{}_{KL}). 
\]
This subtracts
the divergence term
from $\sqrt{-g}R$ and yields the full
Einstein tensor in the bulk.

In analyzing (\ref{action}) it is convenient
to choose the bulk coordinate orthogonal to the
 boundaries: $g_{\mu 5}|_{x^5=0,L}=0$, $0\le\mu\le 3$.
Variation of (\ref{action}) yields again a sum
of a $(4+1)$-dimensional integral and an integral
over the boundary, implying gravitational
equations of motion in the bulk
\be\label{einstein}
R_{MN}=\frac{2\Lambda}{3\mu^3}g_{MN}
\ee
and on the boundary:
\be\label{boundary1}
g^{\lambda\nu}\partial_\mu g_{\lambda\nu}|_{x^5=0,L}
=g^{55}\partial_\mu g_{55}|_{x^5=0,L},
\ee
\be\label{boundary}
\partial_5 g_{\mu\nu}|_{x^5=0}
=-\frac{2}{\mu^3}g_{55}\Big(T_{\mu\nu}^{(1)}
-\frac{1}{3}g_{\mu\nu}g^{\kappa\lambda}T_{\kappa\lambda}^{(1)}\Big),
\ee
\be\label{boundaryb}
\partial_5 g_{\mu\nu}|_{x^5=L}
=\frac{2}{\mu^3}g_{55}\Big(T_{\mu\nu}^{(2)}
-\frac{1}{3}g_{\mu\nu}g^{\kappa\lambda}T_{\kappa\lambda}^{(2)}\Big).
\ee
Here $g_{\mu\nu}$ denotes the tangent components of the
metric tensor on the boundary, and (\ref{boundary1})
 arises from boundary terms $\sim\delta g^{5\mu}$,
while (\ref{boundary}) and (\ref{boundaryb}) arise from boundary 
terms $\sim\delta g^{\mu\nu}$.
No boundary
terms $\sim\delta g^{55}$ appear. 

The energy momentum tensors on the boundary components are
\[
T_{\mu\nu}^{(i)}=-\frac{2}{\sqrt{-g}}\frac{\delta{\cal L}_i}{\delta g^{\mu\nu}},
\]
and as usual in this kind of variational problems, the boundary equations
amount to boundary conditions for the bulk equations of motion.

Eq. (\ref{boundary1}) has two
implications: On the one hand it tells
us that the determinant $-g_{(4)}$ 
 of the metric induced on the boundary
determines the boundary value of $g_{55}$
up to a constant factor, and on the other
hand it ensures invariance of (\ref{action}) under diffeomorphisms
$x^M\to x^M-\epsilon^M(x)$ which leave the boundary invariant:
 $\epsilon^5(x)|_{x^5=0,L}=0$.

To examine the gravitational potential emerging in the
Randall--Sundrum model,
it is useful to reformulate the evolution equations for spatially closed
 $(3+1)$-dimensional boundary universes, i.e. we consider $x^5$ as
a radial coordinate between two spherical shells at radii
 $a\le x^5=r\le b$. Eqs. (\ref{boundary},\ref{boundaryb}) then read
\be\label{boundaryra}
\frac{\partial}{\partial r} g_{\mu\nu}\Big|_{r=a}
=-\frac{2}{\mu^3}g_{55}\Big(T_{\mu\nu}^{(1)}
-\frac{1}{3}g_{\mu\nu}g^{\kappa\lambda}T_{\kappa\lambda}^{(1)}\Big).
\ee
\be\label{boundaryrb}
\frac{\partial}{\partial r} g_{\mu\nu}\Big|_{r=b}
=\frac{2}{\mu^3}g_{55}\Big(T_{\mu\nu}^{(2)}
-\frac{1}{3}g_{\mu\nu}g^{\kappa\lambda}T_{\kappa\lambda}^{(2)}\Big).
\ee

In the Newtonian approximation we consider weakly coupled
 gravitational systems
on time scales much shorter than
the age of the universe and length scales far below the
Hubble radius. In this approximation cosmological
background metrics can very well be approximated by
a local Minkowski background.

The weak field 
approximation $g_{MN}=\eta_{MN}+h_{MN}$ for static
sources $T_{00}^{(i)}=\varrho_i$
on the boundary yields a Neumann type
boundary problem for the gravitational potential $U=-h_{00}/2$:
\be\label{einsteinU}
\Delta U=0,
\ee
\be\label{boundaryUa}
\frac{\partial}{\partial r} U\Big|_{r=a}=\frac{2}{3\mu^3}\varrho_1,
\ee
\be\label{boundaryUb}
\frac{\partial}{\partial r} U\Big|_{r=b}=-\frac{2}{3\mu^3}\varrho_2.
\ee
As a consequence, the gravitational interaction between
matter components on the boundary 
arises through a four-dimensional Greens function
adapted to Neumann boundary conditions:
\be\label{greenU}
U(\vek{r})=\oint_{\partial V} d^3\vek{r}'\Big(G(\vek{r},\vek{r}')
\frac{\partial}{\partial r'} U(\vek{r}')
-U(\vek{r}')
\frac{\partial}{\partial r'} G(\vek{r},\vek{r}')\Big)\Big|_{r'=a}^{r'=b}
\ee
\[
=\langle U\rangle-\frac{2}{3\mu^3}
\int_{r'=a} d^3\vek{r}'
\,G(\vek{r},\vek{r}')\varrho_1(\vek{r}')
-\frac{2}{3\mu^3}
\int_{r'=b} d^3\vek{r}'
\,G(\vek{r},\vek{r}')\varrho_2(\vek{r}').
\]
Here $\langle U\rangle$ is the average value of $U$
on the boundary,  and $d^3\vek{r}'$ is 
the spatial volume element on 
the boundary $\partial V$ of a time slice $V$ of ${\cal V}$.

The Greens function for the Neumann boundary problem
is defined by the requirements
\[
\Delta'G(\vek{r},\vek{r}')=-\delta(\vek{r}-\vek{r}'),
\]
\[
\frac{\partial}{\partial r'} G(\vek{r},\vek{r}')\Big|_{r'=a}
=-\frac{\partial}{\partial r'} G(\vek{r},\vek{r}')\Big|_{r'=b}
=\frac{1}{2\pi^2(a^3+b^3)}
\]
and
we have calculated it for a spatial four-manifold
bounded by two concentric three-spheres:
\be\label{greens}
4\pi^2G(\vek{r},\vek{r}')=
\frac{1}{a^3+b^3}\Bigg(\frac{b^3}{r_>^2}-\frac{a^3}{r_<^2}\Bigg)
+\sum_{l=1}^\infty\Bigg(\frac{r_<^l}{r_>^{l+2}}+\frac{l+2}{l}
\frac{r^lr'^l}{b^{2l+2}-a^{2l+2}}
\ee
\[
+\frac{a^{2l+2}}{b^{2l+2}-a^{2l+2}}
\Big(\frac{r^l}{r'^{l+2}}+\frac{r'^l}{r^{l+2}}
+\frac{l}{l+2}\frac{b^{2l+2}}{r^{l+2}r'^{l+2}}\Big)\Bigg)
\frac{\sin((l+1)\theta)}{\sin\theta}.
\]
Here $\theta$ denotes the angle between the four-dimensional vectors
 $\vek{r}$ and $\vek{r}'$, and like in three-dimensional multipole
expansions $r_<$ is the smaller of the two radii $r$ and $r'$,
while $r_>$ is the larger radius. 

If we choose the three-sphere at $r=a$ as the time slice of our
 $(3+1)$-dimensional universe in this scenario,
the gravitational potential between ordinary matter sources
and probes arises in the limit $r,r'\to a$,
and the distance between source and probe
 within this three-sphere is $d=a\sin\theta$. Up to an irrelevant
constant term the gravitational potential of a mass
 $m$ on the 3-sphere $S^3_a$
of radius $a$ follows from (\ref{greens},\ref{greenU})
\be\label{greensa}
U(\theta)=
-\frac{m}{6\pi^2\mu^3a^2}
\sum_{l=1}^\infty\Bigg(1+\frac{a^{2l+2}}{b^{2l+2}-a^{2l+2}}
\Big(3+\frac{2}{l}+\frac{l}{l+2}\frac{b^{2l+2}}{a^{2l+2}}\Big)\Bigg)
\frac{\sin((l+1)\theta)}{\sin\theta}.
\ee
Here $\theta$ is the angle between the source $m$
of the gravitational field and the point where it is probed.

 A large internal dimension corresponds to $b\gg a$ and yields
\be\label{Ubgga}
U(\theta)|_{b\gg a}=
-\frac{m}{3\pi^2\mu^3a^2}\sum_{l=1}^\infty\frac{l+1}{l+2}
\frac{\sin((l+1)\theta)}{\sin\theta}.
\ee

In the other case of small internal length $L\ll a,b=a+L$ we find
\be\label{UaL}
U(\theta)|_{a,b=a+L\gg L}=
-\frac{m}{3\pi^2\mu^3aL}\sum_{l=1}^\infty\frac{l+1}{l(l+2)}
\frac{\sin((l+1)\theta)}{\sin\theta}.
\ee

 For comparison, the genuine three-dimensional
gravitational potential on a 3-sphere
of radius $a$ is:
\be\label{vs3}
{\cal U}(\theta)=-\frac{m}{4\pi m_{Pl}^2 a}\cot\theta
=
-\frac{2m}{\pi^2 m_{Pl}^2a}\sum_{l=1}^\infty\frac{l}{(2l-1)(2l+1)}
\frac{\sin(2l\theta)}{\sin\theta}, 
\ee
where $m_{Pl}=(8\pi G_N)^{-1/2}$ is the reduced Planck mass
on $S^3_a$.

As expected from a higher-dimensional potential,  
 $U(\theta)|_{b\gg a}$ has a stronger singularity for $\theta\to 0$ 
than the ordinary Newton potential on $S^3_a$. 

The case of small non-periodic extra dimension is more subtle: 
The odd-$l$ modes of $U(\theta)|_{a,b=a+L\gg L}$ are absent 
in the classical inherently three-dimensional 
potential  ${\cal U}(\theta)$, but the even modes agree 
if the naive Kaluza--Klein type relation between $\mu$
and $m_{Pl}$ is augmented by a factor 3: 
\be\label{mumpl} 
3\mu^3 L=m_{Pl}^2.
\ee

Contrary to Kaluza--Klein theory, the relation (\ref{mumpl}) 
in the present theory
eliminates the parameter $L$ completely from the correction 
term to the ordinary Newton potential:
\be\label{correct}
U(\theta)|_{a,b=a+L\gg L}-{\cal U}(\theta)
=
-\frac{m}{4\pi^2 m_{Pl}^2a}\sum_{l=1}^\infty
\frac{2l+1}{l(l+1)}\frac{\sin((2l+1)\theta)}{\sin\theta},
\ee
and therefore the corrections to the Newton potential are not suppressed
by a factor $\exp(-d/L)$ but correspond to an expansion in $d/a$.
In the limit $a\to\infty$, the correction term
corresponds to a renormalization of the
three-dimensional Planck mass, but the functional dependence
on the distance between source and probe
is just that of the three-dimensional Newton potential.

We finally would like to point out that the 4D Poincar\'{e} invariant
 metric of Randall and Sundrum \cite{RS1} complies with the boundary equations
 (\ref{boundary1}--\ref{boundaryb}):

In the present conventions the metric arises from the {\it Ansatz}
\[
g_{\mu\nu}=\exp(-2\sigma(x^5))\eta_{\mu\nu},\quad g_{55}=1
\]
under the assumption of boundary cosmological terms:
\[
{\cal L}_1=-\lambda_1\exp(-4\sigma(0)),
\]
\[
{\cal L}_2=-\lambda_2\exp(-4\sigma(L))
\]
corresponding to boundary energy momentum tensors
\[
T_{\mu\nu}|_{x^5=0}=-\lambda_1g_{\mu\nu}|_{x^5=0},
\]
\[
T_{\mu\nu}|_{x^5=L}=-\lambda_2g_{\mu\nu}|_{x^5=L}.
\]
The Einstein equation in the volume yields again (cf.\ eq.\ (7) in
Ref.\ \cite{RS1})
\[
{\sigma'}^2=-\frac{\Lambda}{6\mu^3},
\]
and the boundary equations imply
\[
\sigma'=\frac{1}{3\mu^3}\lambda_1=-\frac{1}{3\mu^3}\lambda_2,
\]
i.e.\ eq.\ (11) from Ref.\ \cite{RS1} is only rescaled by a factor 4
\[
\Lambda=-\frac{2}{3\mu^3}\lambda_i^2.
\]

We conclude that the criticism of the small-$L$
Randall--Sundrum model was based on an incorrect set of gravitational
evolution equations and not justified. Gravity in the
Randall--Sundrum model is not repulsive, and it has
a correct Newtonian limit. \\[1ex]
{\bf Acknowledgement:} RD thanks Richard Altendorfer for an
interesting discussion.

\end{document}